\documentclass[letterpaper]{article} 
\usepackage{aaai2027}  
\usepackage[hyphens]{url}  
\usepackage{graphicx} 
\usepackage{natbib}  
\usepackage{caption} 
\usepackage{algorithm}
\usepackage{algorithmic}
\usepackage{amsmath}

\usepackage{newfloat}
\usepackage{listings}
\DeclareCaptionStyle{ruled}{labelfont=normalfont,labelsep=colon,strut=off}
\floatstyle{ruled}
\newfloat{listing}{tb}{lst}{}
\floatname{listing}{Listing}

\usepackage{booktabs}
\usepackage{xcolor}

\usepackage{multirow}[c]

\usepackage{multirow}[c]

\title{From Trajectories to Evidence: Auditable Experimental Records \\ for Industrial Research Agents}

\author{
    Zijie Zhuang\equalcontrib,
    Changxin Lao\equalcontrib,
    Pengbo Xu\equalcontrib,
    Hanwen Xu\equalcontrib,
    Ruochen Yang,
    Yingzhi He, \\
    Peng Zhang,
    Jiangxia Cao,
    Yusheng Huang,
    Guohong Mu,
    Jian Liang,
    Ruiming Tang, \\
    Shuang Yang\corresponding,
    Zhaojie Liu,
    Wenwu Ou,
    Kun Gai
}
\affiliations{
    Kuaishou Technology, Beijing, China\\
    \{zhuangzijie, laochangxin, xupengbo03, xuhanwen03, yangruochen, yangshuang08\}@kuaishou.com
}

\begin{document}

\maketitle

\begin{abstract}

Research agents increasingly conduct multi-round machine-learning experiments in industrial recommendation settings and retain the resulting trajectories to guide later decisions.
Yet a completed trajectory is not automatically evidence: generated artifacts may be unsupported or incomplete, executed rounds may be invalid or confounded, and later modifications may obscure earlier findings.
We study \textbf{trajectory-to-evidence conversion}, asking what a completed research process has actually established.
We introduce an evidence-grounded framework that couples bounded verification of consequential artifacts with post-execution claim qualification.
A context-isolated generate--verify--repair process checks artifacts for evidence violations and missing downstream requirements before release.
After execution, validity and attribution checks consolidate evidence across rounds, qualify intervention-level claims as actionable repairs, diagnostic guards, or withheld findings, and preserve admitted claims as auditable records with explicit provenance and applicability boundaries.
A hybrid LLM-assisted controller subsequently applies, defers, or rejects records based on available target evidence.
Record audits characterize which claims survive qualification, while downstream diagnostics identify affirmative applicability judgment as a bottleneck for the tested controller.
Across paper-to-target adaptations, later rounds often improve on the first, while final rounds frequently underperform an earlier best, exposing non-monotonic trajectory evolution.
Candidates produced through the complete workflow also yielded positive online lifts relative to deployed baselines.

\end{abstract}

\section{Introduction}

Research agents increasingly automate iterative machine-learning experimentation.
Given a research objective or paper, an agent may propose an intervention, modify code, construct and execute an experiment, interpret the result, and revise its next action based on the resulting observations~\cite{huang2023mlagentbench,starace2025paperbench,lu2024aiscientist,schmidgall2025agentlaboratory}.
Unlike a single model output, this process produces an extended experimental trajectory containing proposals, code states, execution logs, measurements, failures, and revisions.
Such accumulated trajectories are increasingly retained as 
evolving memory so that prior experience can guide subsequent decisions~\cite{shinn2023reflexion,bytedanceseed2026edgebench,wei2025evomemory}.

An accumulated trajectory records what an agent attempted, but does not directly specify which conclusions are supported by the resulting evidence.
LLM-generated proposals, implementations, and summaries may contain unsupported content or omit information required by downstream steps~\cite{manakul2023selfcheckgpt,farquhar2024semanticentropy}.
Successful execution does not establish that the intended intervention was faithfully realized, the comparison was valid, or the observed change was attributable to the intervention.
Trajectories also mix valid experiments with abandoned hypotheses, retries, failures, and later modifications that may erase earlier gains.
Even a valid result supports a claim only under its observed mechanism, target condition, and evaluation protocol.
Completing a trajectory therefore does not establish which experimental claims are supported strongly enough to be retained.

Existing research agents use feedback and memory primarily to improve subsequent actions.
Self-Evolving Recommendation System~\cite{wang2026selfevolving} coordinates optimization through a shared experiment journal, while NOVA~\cite{liu2026nova} combines semantic verification and historical knowledge to guide architecture search.
Together, they show how experimental feedback and retained history can guide continued optimization, while leaving the qualification of persistent experimental claims largely implicit.

We formulate this problem as \textbf{trajectory-to-evidence conversion}.
Rather than treating each completed round or the final state as experimental knowledge, we consolidate evidence across rounds at the level of experimental claims.
Only findings supported under explicit conditions are preserved as records.
This view separates the evidence established by a trajectory from the sequence of events it contains.

Our framework couples bounded artifact verification with post-execution evidence qualification.
First, a bounded generate--verify--repair process checks consequential artifacts before they affect downstream reasoning or execution.
Second, evidence across executed rounds is consolidated and assessed for execution validity, attribution, and applicability boundaries before supported findings are preserved.
Findings that pass these checks can be retained with the provenance and conditions needed to audit their interpretation.
For downstream use, we evaluate the decisions of a hybrid LLM-assisted controller separately against reference applicability and post-execution outcomes.
We evaluate the framework in industrial recommendation experiments, focusing on trajectory evolution, artifact verification, record construction, downstream applicability, and production case studies.
The contributions of this work are threefold:
\begin{itemize}
    \item We formulate trajectory-to-evidence conversion as the problem of determining which intervention-level claims are supported by evidence distributed across research-agent trajectories.
    \item We introduce a framework that combines bounded verification with cross-round evidence qualification, retaining supported findings with explicit provenance and applicability boundaries.
    \item Through stage-wise analysis, we identify non-monotonic adaptation and unreliable affirmative applicability decisions by the tested LLM-assisted controller as bottlenecks in record construction and downstream use.
\end{itemize}

\section{Related Work}
\label{sec:related-work}

\subsection{Research Agents and Automated Experimentation}

Research agents increasingly automate extended scientific workflows spanning grounding, planning, experimentation, validation, and reporting~\cite{tie2026autoresearch}.
MLAgentBench~\cite{huang2023mlagentbench} evaluates agents that modify code and run iterative machine-learning experiments, while PaperBench~\cite{starace2025paperbench} evaluates the reproduction of published research from paper descriptions.
The AI Scientist~\cite{lu2024aiscientist} and Agent Laboratory~\cite{schmidgall2025agentlaboratory} connect ideation, implementation, experimentation, and writing in broader research pipelines.
These systems establish task-level capabilities for executing and reproducing research workflows, but do not explicitly study which resulting findings warrant retention beyond the current task.

\subsection{Trajectory Memory and Experience Reuse}

Language agents retain prior experience to improve subsequent decisions. Reflexion~\cite{shinn2023reflexion} stores verbal reflections, ExpeL~\cite{zhao2024expel} extracts reusable experience, and Agent Workflow Memory~\cite{wang2025awm} induces workflows from trajectories. Recent studies further examine evolving memory over task streams and long-horizon environments~\cite{wei2025evomemory,bytedanceseed2026edgebench}. In industrial recommendation, agent systems retain experimental history for continued optimization and hypothesis generation~\cite{wang2026selfevolving,wu2026agenticrectune}. NOVA~\cite{liu2026nova} is the closest verification-aware system, combining semantic verification with round-level records and trajectory memory for architecture search.
These approaches use prior trajectories to improve later actions. Systems such as ModelDB and MLflow preserve experimental lineage for traceability and reproducibility~\cite{vartak2016modeldb,zaharia2018mlflow}, but lineage alone does not establish what an experiment supports. Our work instead consolidates evidence across rounds at the intervention-claim level and retains qualified findings with explicit provenance and 
applicability boundaries.

\begin{figure*}[t]
\centering
\includegraphics[
  width=\textwidth,
  height=0.55\textheight,
  keepaspectratio
]{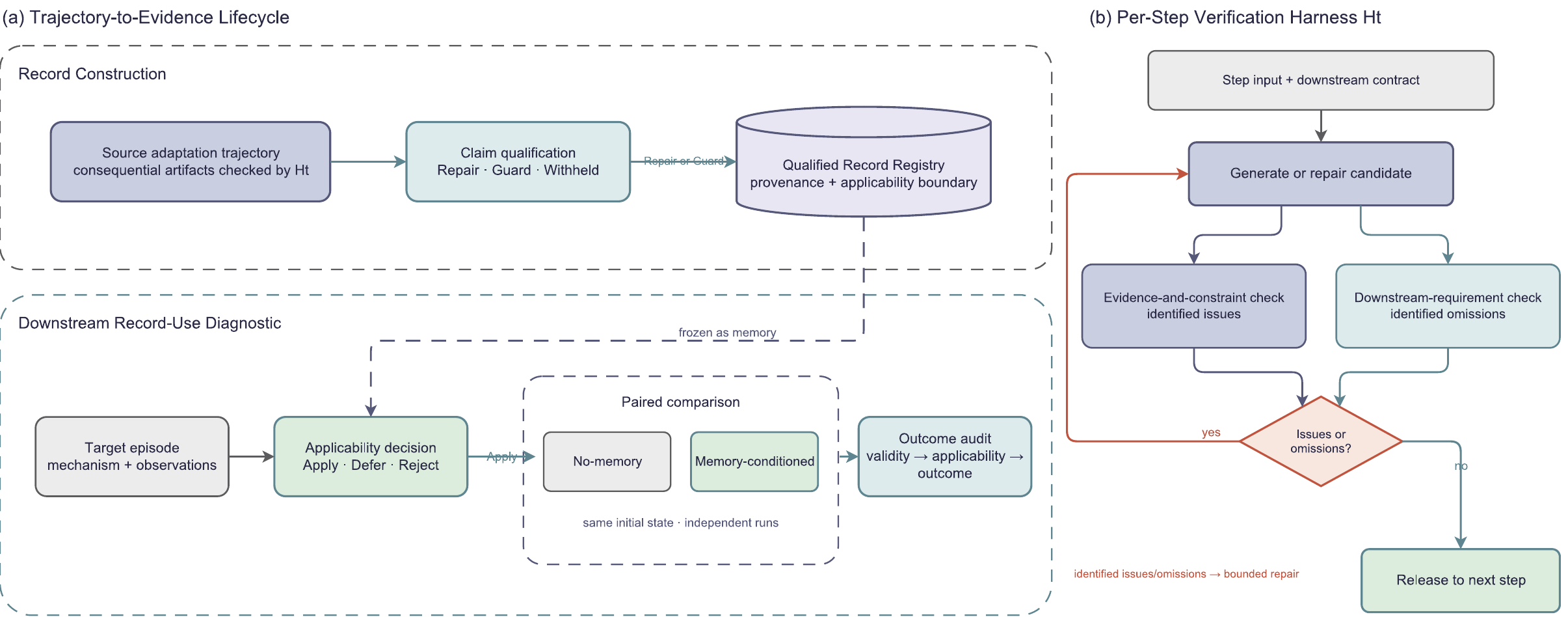}
\caption{
Overview of trajectory-to-evidence conversion.
(a) Verified artifacts from a source adaptation trajectory are qualified as \textsc{Repair}, \textsc{Guard}, or \textsc{Withheld}; only supported \textsc{Repair} and \textsc{Guard} claims enter the registry with provenance and applicability boundaries.
Frozen records are then evaluated in a separately initialized target episode through applicability decisions, paired adaptations for applied Repairs, and outcome audit; applied Guards constrain subsequent planning.
(b) At each consequential handoff, identified issues or omissions trigger bounded repair, and outputs with no remaining issues or omissions are released to the next step.
}
\label{fig:overview}
\end{figure*}

\section{Method}
\label{sec:method}

\subsection{Framework Overview}
\label{sec:method-overview}

Figure~\ref{fig:overview} summarizes three stages: source adaptation, claim qualification, and downstream reuse.
During source adaptation, the agent produces proposals, code changes, experiment specifications, and result summaries.
Before any consequential output moves to the next step, the verification harness checks it against the available evidence, constraints, and downstream requirements.
The producer repairs identified problems, and the harness withholds outputs that still contain unresolved issues after the attempt limit.

After execution, the system groups rounds that address the same intervention and locus and determines which claims their evidence supports and under what observed conditions.
Only qualified claims become records.
We freeze these records before downstream reuse.
Given a target mechanism, target conditions, and available observations, the controller applies, defers, or rejects each candidate record.
An applied Repair becomes a method contract, whereas an applied Guard constrains or redirects subsequent planning.

\subsection{Problem Setting}
\label{sec:problem-formulation}

A paper-to-target adaptation task $\tau$ adapts a paper $p$ to a target setting $c$ while preserving its defining mechanism $m$.
The target setting includes the initial repository, data, evaluation utility, and implementation and deployment constraints.
The defining mechanism identifies the part of the paper that an adaptation must preserve; changes that remove this mechanism do not constitute successful adaptations of $p$.

Starting from an initial implementation, the agent proposes interventions and may dispatch up to $B$ machine-executed experiments.
Each execution produces a code state, runtime observations, and a measured outcome.
The resource limit $B$ counts dispatched machine experiments; it limits machine execution rather than agent-side verification.
The trajectory may contain invalid experiments, retries, and non-monotonic outcomes.
We therefore build claims from evidence across rounds rather than copying completed rounds or the final state.

Our setting contains a set of source adaptation tasks $\mathcal{T}_{\mathrm{src}}=\{\tau_i\}_{i=1}^{n}$ and a target adaptation task $\tau^{\star}$.
Source and target denote roles in the evaluation protocol rather than disjoint paper identities: they may involve the same paper or method, provided that record construction and downstream use occur in separately initialized episodes.
We construct a record registry $\mathcal{M}=\operatorname{ConstructRecords}(\mathcal{T}_{\mathrm{src}})$ from the source trajectories and freeze it before downstream evaluation.
We refer to this frozen record registry as memory when it conditions downstream adaptation.
Only claims qualified as a Repair or Guard yield reusable records; Withheld candidates remain documented but unavailable for reuse.
Each record $r_i$ contains five semantic blocks: source context, intervention, measured outcome, supporting evidence, and applicability boundary.
The context identifies the source mechanism, target condition, and triggering observation, while the intervention includes both the action and its locus.

\subsection{Per-Step Verification Harness}
\label{sec:per-step-harness}

A workflow step is consequential if its output changes a downstream input, determines whether an experiment is dispatched, or contributes to an experimental claim.
For such a step $s_t$, let $x_t$ denote the step input, $E_t$ the evidence available for verification, $z_t$ the output artifact, and $g_{t+1}$ the receiving step's input contract.
The producer samples an initial artifact $z_t^{(0)}\sim G_t(\cdot\mid x_t)$, such as an intervention proposal, code change, experiment specification, or result summary.
Before releasing $z_t$, the harness checks two things.
First, it checks whether the artifact conflicts with the available evidence or constraints.
Second, it checks whether the artifact omits information required by the next step.

The producer, evidence-and-constraint verifier, and requirement reviewer operate in separate contexts.
The producer's private rationale is not shared across these contexts; they communicate only through typed artifacts and explicit issue sets.

\subsubsection{Evidence and Constraint Check}

The evidence-and-constraint verifier receives the step input, the candidate artifact, and the evidence available at that point in the workflow.
It returns a set of localized objections,
\begin{equation}
\mathcal{C}_t^{(k)} =
V_t^{\mathrm{ec}}
\left(x_t,z_t^{(k)},E_t\right),
\end{equation}
where $E_t$ may include the source paper, repository state, tool output, execution logs, or measured results.
An empty set indicates that the verifier found no unsupported statement, inconsistent action, or violated step constraint.
The verifier is instructed to identify concrete problems rather than rewrite the artifact or propose unrelated improvements.

Verification criteria are instantiated according to the artifact type.
Depending on the handoff, these criteria cover consistency with the paper mechanism and observed failure, code semantics and interfaces, experimental comparison and resource constraints, or support for reported conclusions.

\subsubsection{Downstream Requirement Check}

A verifier that reads only the candidate artifact may identify erroneous content while overlooking information that the producer never emitted.
We therefore construct the second check from the next step's information requirements.
The requirement reviewer receives $x_t$ and the downstream contract $g_{t+1}$, but neither the candidate artifact $z_t^{(k)}$ nor the producer's private rationale.
It independently reconstructs the set of information required for the next step:
\begin{equation}
\begin{aligned}
R_t &= V_t^{\mathrm{req}}(x_t,g_{t+1}),\\
\mathcal{O}_t^{(k)} &= \operatorname{Missing}\left(R_t,z_t^{(k)}\right).
\end{aligned}
\end{equation}
The producer receives only the unsatisfied requirements, without access to the reviewer’s hidden reasoning or a replacement artifact.
The reviewer constructs $R_t$ without access to $z_t^{(k)}$.
Afterward, the harness compares $R_t$ with the candidate to return only the missing requirements $\mathcal{O}_t^{(k)}$.
This design keeps the completeness criterion independent of the candidate artifact.

The downstream contract similarly specifies the implementation details, execution interfaces, attribution evidence, or boundary information required by the receiving step.

\subsubsection{Repair and Handoff}

At iteration $k$, the producer receives the current artifact, the objections $\mathcal{C}_t^{(k)}$, and the missing requirements $\mathcal{O}_t^{(k)}$.
If both sets are empty, the artifact is released to $s_{t+1}$.
Otherwise, the producer revises the artifact, and the revised artifact is independently checked again by both verifiers.
This generate--verify--repair process continues for at most $K_t$ repair attempts after the initial generation.
If no candidate passes within this limit, the harness stops the current branch, rolls back to a previously validated state, or requests escalation.
The artifact is released when both checkers report no unresolved issues or omissions.
The limit $K_t$ controls agent-side verification effort.

\subsection{Experimental Claim Qualification}
\label{sec:evidence-admission}

Passing the harness establishes artifact readiness for downstream execution, whereas intervention effectiveness is assessed from the resulting experiment.
After the pre-execution artifacts pass, the system dispatches $e_b$; its implementation, outcome, interpretation, and summary re-enter the harness as experimental evidence.

We distinguish execution validity from claim attribution.
An experiment is execution-valid when the implementation realizes the proposed intervention, the prescribed run completes, the comparison follows the evaluation protocol, and the paper's defining mechanism is preserved.
A negative or null result may remain a valid execution, but it supports a persistent Repair or Guard only when the observed effect is attributable and explicitly bounded to the tested conditions.

Let $\mathcal{G}_{\mathrm{exec}}$ denote the implementation-fidelity, execution, evaluation, and mechanism-preservation gates.
An executed round is valid only when all four gates pass:
\begin{equation}
\operatorname{ExecutionValid}(e_b)
=
\bigwedge_{g\in\mathcal{G}_{\mathrm{exec}}} I_g(e_b).
\end{equation}
Implementation fidelity checks that the code realizes the proposed intervention, and execution requires the prescribed run to complete.
Evaluation checks the comparison protocol, and mechanism preservation checks that the adaptation retains the paper's defining mechanism.
These gates may combine deterministic tests, execution evidence, controlled comparisons, and structured assessments.
Runs that fail an execution gate are retained as operational diagnostics but excluded from persistent claims.
Attribution is assessed separately during claim qualification; an execution-valid run with insufficient attribution is likewise retained only as a diagnostic.

The qualification step groups rounds that address the same intervention and locus, then reads their verified code changes, execution logs, measurements, and parent comparisons.
Related execution-valid rounds may provide primary, corroborating, or boundary evidence; invalid, confounded, duplicated, or superseded rounds are excluded.
Based on this evidence, the qualification step labels the candidate as an actionable Repair, a diagnostic Guard, or Withheld.

A Repair requires execution-valid and attributable evidence for an implementable intervention, together with an explicit trigger and applicability boundary.
A negative or null result yields a Guard only when it establishes an attributable failure mode under explicit observed conditions.
The system withholds a candidate when the execution-valid rounds do not support a consistent claim, attribution is insufficient, or the trigger or boundary cannot be established.

Each Repair or Guard becomes a record $r_i$.
The record states why the intervention was attempted and where it acts, and links to the implementation diff, execution status, protocol, measurements, and attribution checks.
It also records the observed conditions that limit reuse.
Each qualified record therefore represents a source-supported claim with an explicit applicability boundary.

\subsection{Downstream Use of Experimental Records}
\label{sec:memory-transfer}

For a target adaptation task, the controller queries the frozen record registry $\mathcal{M}$ and forms
\begin{equation}
\begin{aligned}
q_t^{\star}&=\left\langle m^{\star},c^{\star},o_t^{\star}\right\rangle,\\
d(r_i,q_t^{\star})&\in
\{\mathrm{Apply},\mathrm{Defer},\mathrm{Reject}\},
\end{aligned}
\end{equation}
from the target method's defining mechanism, target conditions, and available target observations.
Retrieval first identifies records with potentially compatible mechanisms and conditions, after which the adaptation controller assigns each candidate record a reuse decision.
The controller combines deterministic checks over structured record and target fields with an LLM-based assessment of mechanism, condition, and intervention compatibility.

The controller assigns \textsc{Apply} when its assessment supports mechanism and locus compatibility, required triggers, boundary compliance, and preservation of the target mechanism.
It assigns \textsc{Defer} when the intervention locus may be compatible but a required trigger, boundary fact, or observable remains missing or ambiguous.
It assigns \textsc{Reject} when the evidence indicates an incompatible mechanism or intervention locus, a boundary violation, or an inability to preserve the target method's defining mechanism.
An \textsc{Apply} decision activates a target-specific experiment based on the available record and observations; its scientific outcome is determined after execution.

An applied Repair becomes an implementable method contract specifying its intervention locus, connection to the target method, required interfaces, configuration, and mechanism-preservation constraints.
The contract is evaluated by the same per-step harness as a newly generated proposal before execution.
An applied Guard instead becomes a planning constraint or diagnostic directive and does not by itself dispatch an intervention.

\section{Experiments}
\label{sec:experiments}

\subsection{Experimental Setup}
\label{sec:experimental-setup}

\paragraph{Agent model and verification setup.}
All LLM components, including the controller's semantic assessment, use GLM-5.2~\cite{glm5team2026glm5} self-hosted on our GPU infrastructure, with separate contexts for generation, evidence checking, and requirement review.
Because GPU-backed training and evaluation dominate cost and latency in our setting, we budget dispatched machine experiments rather than inference tokens.
Agent-side verification is controlled by the fixed retry limit $K_t$ and runs before dispatch to reduce the risk that invalid artifacts consume the experiment budget.

\paragraph{Evaluation scope.}
We align each evaluation with its corresponding stage-level question: whether consequential artifacts satisfy downstream requirements, whether persistent records are supported and traceable, and how record use relates to reference applicability and observed outcomes.
We report performance differences for valid, mechanism-preserving executions.
We reserve intervention-level attribution for pairs in which code inspection confirms equivalent base implementations and isolates memory as the intervention.
Other matched executions are interpreted at the task level.
Accordingly, we report stage-level effects and candidate-level production outcomes rather than an aggregate causal estimate of the complete workflow.

\subsection{Non-Monotonic Adaptation Trajectories}
\label{sec:iterative-adaptation}

\begin{figure*}[t]
\centering
\includegraphics[
  width=\textwidth,
  height=0.6\textheight,
  keepaspectratio
]{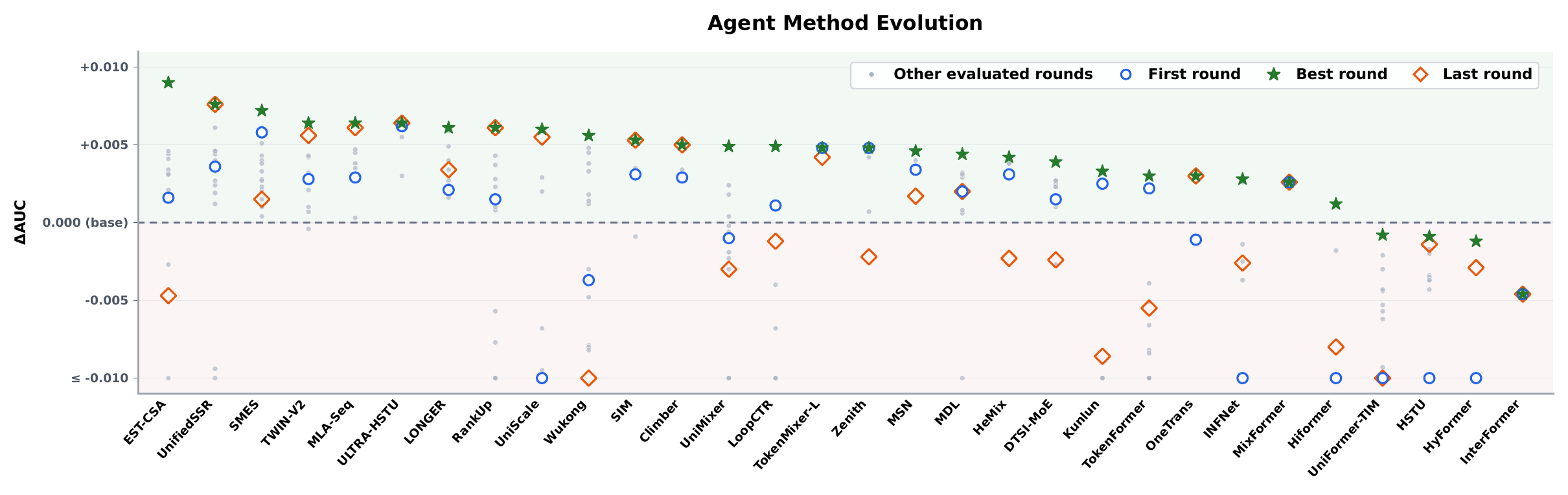}
\caption{
\textbf{Iterative adaptation is effective but non-monotonic.}
Starting from the same industrial RankMixer baseline, our agent adapts methods from 30 papers through multiple experimental rounds.
For 26 of the 30 methods, at least one subsequent round outperforms the first; however, for 22 methods, the final round performs worse than an earlier best.
These results show that iterative adaptation often discovers a better observed state but does not preserve gains monotonically.
}
\label{fig:main_exp}
\end{figure*}

We use a production implementation of RankMixer~\cite{zhu2025rankmixer} as the shared baseline and adapt modules from 30 source papers to it.
For each paper, the research agent conducts multiple rounds of intervention, implementation, execution, and evaluation.
We retain the first, best, and last valid round and report their metric differences from the production baseline.
Figure~\ref{fig:main_exp} summarizes the resulting 30 industrial paper-to-target adaptations.
For 26 methods, at least one later valid round outperforms the first, showing that continued experimentation often discovers a better observed state. 
However, these gains are not preserved monotonically: for 22 methods, the last round falls below an earlier best, and later modifications sometimes partially or completely erase a previously observed gain.

Thus, the final state of an adaptation trajectory is not necessarily its most useful experimental knowledge. 
This observation motivates constructing memory from validated intermediate interventions and outcomes instead of copying the trajectory endpoint.

\subsection{Bounded Self-Verification}
\label{sec:verification-results}

Table~\ref{tab:verification-evidence} summarizes two context-isolated verification handoffs.
At the paper-to-proposal handoff, generation, evidence-and-constraint verification, and requirement review use mutually isolated contexts of the same GLM-5.2 model.
Of the 15 paper-reproduction bundles, 11 pass on the first attempt; allowing up to three verification-and-repair retries increases gate acceptance from 73.3\% to 100\%, without manual correction.
At the proposal-to-code handoff, the loop increases implementation completeness from 92\% to 97\% and observability from 67\% to 100\%, while alignment changes only marginally from 50\% to 52\%.

\begin{table}[t]
\centering
\footnotesize
\setlength{\tabcolsep}{5pt}
\begin{tabular}{@{}lcc@{}}
\toprule
\textbf{Criterion} & \textbf{Baseline (\%)} & \textbf{Loop (\%)} \\
\midrule
Proposal gate acceptance & 73.3 & 100 \\
Code completeness & 92 & 97 \\
Code alignment & 50 & 52 \\
Code observability & 67 & 100 \\
\bottomrule
\end{tabular}
\caption{Context-isolated self-verification at two representative consequential handoffs. Proposal results compare the first attempt with the accepted output; code results compare artifacts produced without and with the loop.}
\label{tab:verification-evidence}
\end{table}

Overall, the observed gains come mainly from improved completeness and observability, rather than tighter alignment with the proposal.
Because generation and verification share the same model parameters, context isolation prevents direct rationale sharing but may leave correlated blind spots.
Table~\ref{tab:verification-evidence} evaluates loop behavior at two handoffs, whereas Table~\ref{tab:verification-audit} independently audits the final accepted bundles.
All audited evidence spans are traceable and all core-mechanism groups have source support, although coverage of the source papers' Method sections remains incomplete.
Together, these results indicate that same-model verification provides a measurable but incomplete correction signal.

\begin{table}[t]
\centering
\footnotesize
\begin{tabular}{@{}lr@{}}
\toprule
\textbf{Criterion} & \textbf{Pass / Total} \\
\midrule
Source-traceable spans
    & 479/479 (100\%) \\
Source-supported mechanisms
    & 107/107 (100\%) \\
Covered Method sections
    & 129/139 (92.8\%) \\
Covered Method blocks
    & 458/524 (87.4\%) \\
\bottomrule
\end{tabular}
\caption{Post-hoc source-support and coverage audit of the final paper-reproduction bundles.}
\label{tab:verification-audit}
\end{table}

\subsection{Automated Reproduction on a Public Benchmark}
\label{sec:public-reproduction}

To assess whether the agent pipeline can translate paper specifications into executable experiments, we apply the adaptation pipeline to five models on the Amazon Electronics subset of the Amazon product-review dataset~\cite{mcauley2015image}.
Table~\ref{tab:public-reproduction} shows that the reproduced results preserve the reported ordering (Spearman $\rho=1.0$), and every reproduced non-FM model remains above FM.

\begin{table}[t]
\centering
\small
\setlength{\tabcolsep}{1.5pt}
\begin{tabular}{lccccc}
\toprule
\multirow{2.5}{*}{\textbf{Model}} & \multirow{2.5}{*}{Rank} & \multicolumn{2}{c}{AUC}
& \multicolumn{2}{c}{$\Delta_{\mathrm{FM}}$} \\
\cmidrule(lr){3-4}\cmidrule(lr){5-6}
 &
& Reported & Reproduced
& Reported & Reproduced \\
\midrule
InterFormer & 1 & 0.8865 & 0.8522 & +0.0380 & +0.0548 \\
DIN         & 2 & 0.8848 & 0.8470 & +0.0363 & +0.0496 \\
DHEN        & 3 & 0.8790 & 0.8436 & +0.0305 & +0.0462 \\
Wukong      & 4 & 0.8765 & 0.8413 & +0.0280 & +0.0439 \\
FM          & 5 & 0.8485 & 0.7974 & --      & --      \\
\bottomrule
\end{tabular}
\caption{Comparison of reported and agent-produced results on Amazon Electronics.}
\label{tab:public-reproduction}
\end{table}

\subsection{Experimental Record Audit}
\label{sec:record-audit}

We audit the conversion from completed trajectories to persistent experimental records.
The pipeline consolidates 14 candidate claims: eight are qualified as actionable Repairs, one as a diagnostic Guard, and five are Withheld.
Only the Repairs and Guard produce records, yielding a frozen registry of nine records, as summarized in Table~\ref{tab:record-qualification}.

\begin{table}[t]
\centering
\footnotesize
\setlength{\tabcolsep}{3pt}
\begin{tabular}{@{}lrrp{1.55in}@{}}
\toprule
Type & Claims & Records & Main reason \\
\midrule
Repair & 8 & 8 & Attributable intervention with an explicit trigger and boundary \\
Guard & 1 & 1 & Attributable negative evidence supporting a bounded failure mode \\
Withheld & 5 & 0 & Confounded or insufficient evidence, an unresolved trigger, or an outcome inconsistent with the claim \\
\midrule
Total & 14 & 9 & -- \\
\bottomrule
\end{tabular}
\caption{Qualification outcomes for the 14 candidate claims.}
\label{tab:record-qualification}
\end{table}

All nine records contain the five semantic blocks defined in our method and specify an explicit trigger and boundary.
The eight Repairs can be compiled into executable method contracts, while the Guard encodes a downstream planning constraint.
All 34 referenced source artifacts match their archived SHA-256 values, and all five Withheld candidates are excluded from reusable entries.
The registry snapshot was frozen before the downstream record-use diagnostics.

Table~\ref{tab:record-cases} illustrates how evidence structure, rather than outcome sign alone, determines qualification: an attributable intervention supports a Repair, a repeated and bounded failure supports a Guard, and an unresolved trigger remains Withheld.

\begin{table*}[t]
\centering
\footnotesize
\setlength{\tabcolsep}{4pt}
\begin{tabular}{@{}
p{0.55in}
p{1.50in}
p{2.35in}
p{1.85in}
@{}}
\toprule
\textbf{Type}
& \textbf{Representative case}
& \textbf{Evidence}
& \textbf{Qualification decision} \\
\midrule

Repair
& EST: pre-head RMSNorm
& An ablation showed that removing RMSNorm reduced mean AUC from 0.6526 to
0.5433, raised the head-input standard deviation to 5340, and collapsed
three of four datasets to AUC 0.5. Similar backbone diagnostics localized
the failure to the prediction interface.
& Repair: normalize an observably unstable final representation before the head.
The supported claim is localized to the prediction interface. \\

\addlinespace

Guard
& RankMixer: inactive sparse gate
& Across three successive repair attempts on four datasets, gate activity
remained near zero. Neither connecting the gate to the task path nor
reducing the L1 penalty tenfold restored it; the latter also reduced AUC by
0.0226.
& Guard: avoid further unguided gate tuning and inspect gradient reachability, initialization, and sparsity pressure.
No Repair is admitted. \\

\addlinespace

Withheld
& TokenMixer: $W_{\mathrm{down}}$ initialization
& Reducing the initialization scale from 0.01 to 0.001 did not restore the
intended residual signal and produced only a negligible AUC change.
& Withheld: the trigger remains unresolved, with no supported repair or
reusable boundary. \\

\bottomrule
\end{tabular}
\caption{Representative outcomes of experimental claim qualification.}
\label{tab:record-cases}
\end{table*}

\subsection{Memory-Conditioned Adaptation}
\label{sec:memory-conditioned-results}

We conduct a paired diagnostic of how a frozen source memory changes the initial outcome of target adaptation.
The diagnostic set contains eight target--memory pairs covering seven target methods. 
For each pair, we independently execute a no-memory run and a memory-conditioned run sharing target paper, initial baseline, data, evaluation protocol, random seed, machine-experiment allowance, and defining mechanism.
The no-memory run receives only the target Method, whereas the memory-conditioned run adds one source-derived intervention. 
We fix the record before launching either arm, and outcomes from the pair neither modify that record nor enter the frozen registry.
These interventions cover normalization, expert density, optimization schedules, token budgets, feature sharing, and mixer depth.

Before execution, an outcome-blind reference audit labels each pair from the target mechanism, observed conditions, and recorded boundary.
To separate applicability from controller error, we execute all five reference-\textsc{Apply} pairs and three reference-\textsc{Defer}/\textsc{Reject} pairs with the gate bypassed.
All eight pass the execution gates and are task-level comparable under the protocol above.
After execution, we separately classify the task-level outcome as beneficial, neutral, or harmful.

\begin{table*}[t]
\centering
\small
\setlength{\tabcolsep}{4pt}
\begin{tabular*}{\textwidth}{
  @{\extracolsep{\fill}}
  p{1.0in}
  p{1.6in}
  p{1.05in}
  r
  l
  @{}
}
\toprule
Target & Transferred intervention & Reference applicability
& $\Delta\mathrm{AUC}$ & Outcome \\
\midrule
TokenFormer & Branch-output RMSNorm
    & \textsc{Reject} & -0.0227 & Harmful \\
RankMixer & Sparse-to-dense experts
    & \textsc{Reject} & +0.0003 & Neutral \\
InterFormer & Residual-output RMSNorm
    & \textsc{Defer} & -0.0032 & Harmful \\
MixFormer & Pre-head RMSNorm
    & \textsc{Apply} & +0.0030 & Beneficial \\
TokenFormer & Warmup + higher LR
    & \textsc{Apply} & +0.0034 & Beneficial \\
OneTrans & Per-feature token cap
    & \textsc{Apply} & +0.0101 & Beneficial \\
HiFormer & FFN shared across features
    & \textsc{Apply} & -0.0187 & Harmful \\
HeMix & Single mixer block
    & \textsc{Apply} & -0.0007 & Neutral \\
\bottomrule
\end{tabular*}
\caption{Memory-conditioned adaptation results.
$\Delta\mathrm{AUC}$ is the memory-conditioned AUC minus the no-memory AUC.
Values with $\Delta\mathrm{AUC}>0.001$, $\Delta\mathrm{AUC}<-0.001$, and $|\Delta\mathrm{AUC}|\leq0.001$ are labeled Beneficial, Harmful, and Neutral, respectively.}
\label{tab:memory-conditioned}
\end{table*}

Table~\ref{tab:memory-conditioned} shows that memory reuse is conditional.
Of the five \textsc{Apply} pairs, three are beneficial, one is neutral, and one is harmful. 
Of the three manually dispatched \textsc{Defer} or \textsc{Reject} pairs, two are harmful and one is neutral. 
Reference applicability filters plausible reuse, whereas execution determines the target-specific outcome.

\subsection{Controller Error Analysis}
\label{sec:controller-diagnostic}

\begin{table}[t]
\centering
\small
\begin{tabular}{lccc}
\toprule
Reference & Apply & Defer & Reject \\
\midrule
Apply  & 1 & 1 & 0 \\
Defer  & 1 & 9 & 1 \\
Reject & 2& 16 & 33 \\
\bottomrule
\end{tabular}
\caption{Controller decisions on a frozen diagnostic set. Columns are
controller outputs.}
\label{tab:applicability-audit}
\end{table}

We separately characterize how the adaptation controller maps available target evidence to \textsc{Apply}, \textsc{Defer}, or \textsc{Reject} on a frozen diagnostic set of 64 target--record pairs.
The reference decisions follow the same outcome-blind applicability protocol used in Table~\ref{tab:memory-conditioned}.
Table~\ref{tab:applicability-audit} reports the resulting confusion counts.

The hybrid controller issued four \textsc{Apply} decisions, only one of which matched the reference, yielding an affirmative precision of 25\%; the remaining three corresponded to one reference-\textsc{Defer} and two reference-\textsc{Reject} cases.
Its overall accuracy was 67.2\%, below the 79.7\% obtained by an always-\textsc{Reject} rule on this class-imbalanced set.
However, always rejecting yields zero \textsc{Apply} recall and dispatches no record-conditioned experiments, so it cannot generate target-local evidence from reuse.
Aggregate accuracy therefore rewards inactivity and is not an adequate operational objective in this setting.
The observed false affirmatives show that the tested controller has not yet achieved a reliable balance between conservative filtering and experimental progress.
We therefore treat an \textsc{Apply} decision as a hypothesis for execution rather than a confirmation of applicability.

\subsection{Production Case Studies}
\label{sec:industrial-validation}

\begin{table}[t]
\centering
\small
\begin{tabular}{lcc}
\toprule
Scenario & Business metric & Online gain \\
\midrule
Live streaming & Page duration & +0.75\% \\
User growth & Net growth utility & +6.34\% \\
\bottomrule
\end{tabular}
\caption{Observed online lifts of two candidates produced through the complete workflow over production baselines.
}
\label{tab:industrial-results}
\end{table}

Although the previous diagnostics expose limitations in record use, they do not prevent the workflow from producing deployable candidates.
Each candidate originated from a paper-derived intervention and passed artifact verification, machine evaluation, and claim qualification before online testing.
These cases connect trajectory-to-evidence conversion with model evaluation under production constraints.
As shown in Table~\ref{tab:industrial-results}, two candidates from the complete workflow yielded positive online lifts over frozen production baselines in real-world A/B tests.

\section{Limitations}

The production case studies establish candidate-level online performance against frozen baselines rather than framework-level attribution, which would require a parallel plain-agent or no-memory arm.
Because stochastic upstream changes alter downstream artifacts, attribution is interpreted at the task level except for code-inspected equivalent implementations.
The record-use diagnostic contains only eight single-run pairs, and the controller evaluation is class-imbalanced, limiting conclusions about generalization.

\section{Conclusion}

We present a framework that verifies research-agent experiments and converts supported findings into auditable, boundary-aware records.
Across paper-to-target adaptations, final states often lose earlier gains, and several candidate claims lack sufficient evidence for retention.
Downstream diagnostics identify affirmative applicability judgment by the tested controller as a remaining bottleneck.
Separately, two candidates produced through the complete workflow yielded positive online lifts over frozen production baselines.
Overall, our results support treating a trajectory as a source of candidate evidence requiring explicit verification, rather than as experimental knowledge by itself.

\bibliography{aaai2027}

\end{document}